\documentstyle[graphicx]{mn}
\newcommand{\ls}
 {\mathrel{\hbox{\rlap{\hbox{\lower4pt\hbox{$\sim$}}}\hbox{$<$}}}}
\newcommand{\gs}
 {\mathrel{\hbox{\rlap{\hbox{\lower4pt\hbox{$\sim$}}}\hbox{$>$}}}}
\newcommand{\degg}{\hbox{$^\circ$}}

\newcommand{\et}{et al.\ }
\newcommand{\rosat}{{\it ROSAT}}

\newcommand{\asca}{{\it ASCA}}
\newcommand{\xte}{{\it RXTE}}
\newcommand{\sax}{{\it BeppoSAX}}
\newcommand{\xmm}{{\it XMM}}
\title[Ionised Disc in Ark~564]
	{ Evidence for an ionised disc in the narrow-line Seyfert 1
galaxy Ark~564 }
\author[Vaughan \et]
        {S.\ Vaughan$^{1}$, K. A.\ Pounds$^{1}$, J.\ Reeves$^{1}$, 
R.\ Warwick$^{1}$, R.\ Edelson$^{1,2}$\\
$^1$X-Ray Astronomy Group; Department of Physics and Astronomy;
Leicester University; Leicester LE1 7RH; U.K.\\
$^2$Department of Physics and Astronomy; University of California,
Los Angeles; Los Angeles, CA 90095-1562; U.S.A.\\
}

\date{Accepted 1999 July 30. Received 1999 July 13; in original form 1999 May
27} 
\pagerange{\pageref{firstpage}--\pageref{lastpage}}
\pubyear{1999}
\begin{document}
\maketitle
\label{firstpage}

\begin{abstract}
We present simultaneous ASCA and RXTE observations of Ark~564, the
brightest known `narrow-line' Seyfert~1 in the 2--10~keV band. The measured
X-ray spectrum is dominated by a steep ($ \Gamma \approx 2.7 $) power-law
continuum extending to at least 20 keV, with imprinted Fe K-line and -edge
features and an additional `soft excess'  below $\sim 1.5 $ keV.
The energy of the iron K-edge indicates the presence of highly ionised
material, which we identify in terms of reflection from a strongly
irradiated accretion disc. The high reflectivity of this putative disc,
together with its strong intrinsic O~\textsc{viii} Ly-$\alpha$ and 
O~\textsc{viii} 
recombination emission, can also explain much of the observed soft excess
flux. Furthermore, the same spectral model also provides a reasonable
match to the very steep 0.1--2 keV spectrum deduced from ROSAT data.
The source is much more rapidly variable than `normal' Seyfert~1s of
comparable luminosity, increasing by a factor of $\sim$50\% in
1.6~hours, with no measurable lag between the 0.5--2~keV and 3--12~keV bands,
consistent with much of the soft excess flux arising from reprocessing of
the primary power-law component in the inner region of the accretion
disc. 
We note, finally, that if the unusually steep power-law component
is a result of Compton cooling of a disc corona by an intense
soft photon flux, then the implication is that the bulk of these soft
photons lie in the unobserved extreme ultraviolet.
\end{abstract}

\begin{keywords}
galaxies: active -- galaxies: individual: Ark~564 -- X-rays: galaxies
\end{keywords}

\section{Introduction }

Narrow-line Seyfert~1 galaxies (NLS1), defined as having H$\beta$
FWHM~$\leq$~2000~km/s (Osterbrock \& Pogge 1985), possess distinctive
X-ray properties that set them apart from `normal,' broad-line
Seyfert~1 (BLS1) galaxies.  \rosat\ observations have shown that
NLS1s exhibit rapid variability and steep spectra in the soft X-ray band 
(Boller \et 1996) with more recent \asca\ measurements
revealing that this anomalous spectral steepness often extends into the
2--10~keV band (Brandt \et 1997; Vaughan \et 1999). 

It seems likely that the distinctive observational properties of this
sub-class of AGN relate to some fundamental physical parameter common to
all NLS1. By analogy with the `high state' properties of Galactic
black-hole candidates (GBHCs), it has been suggested that the fundamental
`driver' in NLS1 is the central black hole accreting at or above the
Eddington limit (Pounds \et 1995; Laor \et 1997).  
One consequence of a high accretion rate might be that most of the
power in NLS1 is liberated in a hot, photoionised accretion disc rather
than a disc corona (Ross \et 1992). Thermal emission from such a hot disc
could then explain the strong soft excess flux and steep hard X-ray
continuum often seen in NLS1s, as the copious soft photons will
Compton cool the disc corona resulting in a steepening of the spectrum 
(Pounds \et 1995). 

The X-ray spectrum should
also bear the signature of Compton `reflection' from the disc
surface, which is expected to be highly ionised in high accretion rate objects
(e.g., Matt \et 1993). The form of the reflection features,
particularly the iron K line and absorption edge and the form of the soft
X-ray continuum, should therefore differ in NLS1s from BLS1s if indeed
NLS1s are accreting at a higher rate. 
Tentative support for this view has recently been presented by Comastri
\et (1998a) and Turner \et (1999) who find evidence for an emission
line near 7~keV, consistent with K$\alpha$ emission in hydrogen-like
iron in the NLS1 Ton~S180.

Ark~564 is the brightest known NLS1 in the 2--10 keV band and
hence can be considered an ideal object in which to study the spectral
features described above. 
\rosat\ data revealed a complex spectrum in the soft
X-ray band, well fitted with either a power law and strong soft
excess, or a (steeper) power law and an absorption edge at 1.2~keV
(Brandt \et 1994). Vaughan \et
(1999) in their study of the \asca\ spectra of 22 NLS1 find evidence in
Ark 564
for an emission feature at $\sim$1~keV or a broad
absorption feature at 1.2~keV. In the present paper we perform a more 
detailed analysis of these
\asca\ data together with simultaneous \xte\ observations of Ark~564, to
provide the first study of a NLS1 out to 20~keV. 
All line and edge energies derived from the spectral fitting are given
in the rest frame of the source and errors are quoted at the 90\%
confidence level 
unless stated otherwise.

\section{Observations and data reduction}

\subsection{The ASCA observations}

Ark~564 was observed by \asca\ on 23--25 December 1996 for a duration of
103~ksec. 
After applying standard screening
criteria, the total `good' exposure time reduced to 47~ksec. In both
pairs of SIS and GIS instruments counts were accumulated from a 4\arcmin\
circular aperture centred on the source position, with background
estimated from source free regions at similar off-axis angles. The
derived pulse height spectra were binned to give at least 20 counts per
spectral channel. There appear to be significant calibration problems in
these \asca\ spectra at soft energies; the SIS and GIS spectra diverge at
$\sim 1$~keV, differing by 30\% at 0.8~keV, and the two SIS spectra
diverge from each other at lower energies. This is most likely due to
radiation damage to the SIS CCDs.
The response of the GIS detectors is not thought to be 
time-dependent, but their sensitivity is poor below
1~keV. In order to minimize the effect of these
uncertainties in the spectral analysis presented here, the SIS data below
1.0~keV have been ignored, as have the GIS data below 0.8~keV.

We note that the SIS lower-level discriminator setting changed during the
observation. In order to assess the impact on the spectral analysis, we
generated separate SIS spectra and response matrices for each
discriminator setting. The difference between the spectra obtained with
the different discriminator settings was not significant and was much
smaller than the difference between the SIS-0 and SIS-1 spectra. We
therefore did not distinguish between data gathered with different
discriminator settings in the following analysis.

Source and background light curves were extracted in 128s bins, from
both SIS detectors, in the soft (0.5--2~keV) band. (Note that a softer
band was used for the light curves because the spectral calibration
problems above have a much smaller effect on temporal analysis.) The
two background subtracted light curves were combined to increase
signal/noise and the resulting light curve was binned by orbit (5760s)
for the temporal analysis.

\subsection{The RXTE observations} 

Ark~564 was observed by \xte\ simultaneously with \asca\ for a
duration of 95~ksec. 
Data from the top (most sensitive) layer of the 
PCU array was extracted using the
\textsc{rex} reduction script supplied by NASA/GSFC.
Poor quality data were excluded on the basis of the
following acceptance criteria: the satellite is out of the South Atlantic
Anomaly; Earth elevation angle $\geq$~10\degg; offset from optical
position of Ark~564 $\leq$~0.02\degg; and
\textsc{electron-0}$\leq$~0.1. This last criterion removes data
with high anti-coincidence rate in the PCUs. The total `good' exposure
time selected was 42~ksec. 
Data were collected from PCUs 0,1 and 2 and
the background was estimated using the L7--240 model.

As with the \asca\ data, the background-subtracted pulse height spectrum
from the PCU array was binned to give at least 20 counts per
channel. 
In the subsequent spectral fitting the
\xte\ data below 2.5~keV have been ignored in order to avoid
calibration uncertainties, as well as above 20~keV where the
signal is almost entirely background. The hard band (3--12~keV) light
curve was extracted in 16s bins and, as with the \asca\ light curve,
was rebinned by orbit.

\section{Spectral analysis}

\begin{figure}
 \begin{center}
\rotatebox{-90}{\includegraphics[width=5.5cm]{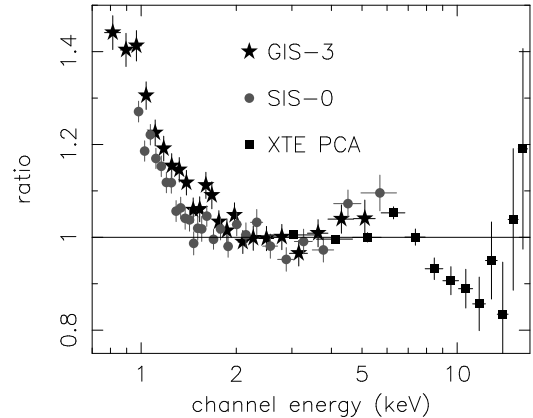}}
 \end{center}
 \caption {The data/model ratio from a simple power law fit to the
combined \xte\ and \asca\ data. The data were fitted in the hard X-ray 
band using model 1 of Table 1. Note the large deviations in the
5--9~keV range and the soft excess below 1.5~keV.}
\end{figure}

The background subtracted spectra from \asca\ and \xte\ were fitted using
the \textsc{xspec v10.0} software package. After examining the
spectral fits separately to 
check for cross-calibration problems, we then fitted the spectra from
both satellites simultaneously but with the relative normalisations free
to vary. The \xte\ spectrum overestimates the flux in the overlapping
2.5--10~keV band compared with \asca, but there are no
systematic differences in the shape of the spectrum.
The fitting of a simple power-law model to the combined \asca\ and \xte\
spectra revealed strong residuals in the 5--9~keV band and a strong soft
excess below 1.5~keV (as in Fig.~1).

\begin{table*}
\centering
\caption{Results of simultaneous fits to the \asca\ and \xte\ data
in the 2.5--20~keV range. The columns provide the following information:
(1) the type of spectral model (as defined in the text); 
(2) the power-law photon index; 
(3) and (4) the iron line energy (keV) and equivalent width (eV) respectively; 
(5) the iron edge energy (keV) or the ionisation parameter; 
(6) the edge depth, the column density of absorbing gas ($10^{22}$~cm$^{-2}$) or 
the reflection strength;
(7) the best-fit $\chi^{2}$ and number of degrees of freedom in the fitting.} 
\begin{tabular}{@{}lcccccc@{}}                 
Model & $\Gamma$ & E$_{line}$  & EW  & E$_{edge}$/$\xi$ & 
$\tau$/N$_{H}$/R    & $\chi^{2}$/dof \\   
 (1)    &  (2)     & (3)  & (4) & (5) & (6) & (7)  \\
\hline
1. PL & $2.53\pm0.02$ & -- & -- & -- & --  & 1087/1062  \\
2. PL+LINE & $2.56\pm0.02$   & $6.43^{+0.26}_{-0.12}$ & $95^{+34}_{-18}$ & -- & 
--  & 1047/1060 \\
3. PL+LINE+EDGE & $2.54\pm0.02$ & $6.4^{f}$ & $95\pm28$ & $7.1^{f}$ & $<0.1$  & 
1045/1060  \\
4. PL+LINE+EDGE & $2.51\pm0.02$ & $6.7^{f}$ & $74\pm30$ & $8.76^{f}$ & 
$0.22\pm0.07$  & 1024/1060  \\
5. PL+LINE+EDGE & $2.51\pm0.02$ & $6.42^{+0.35}_{-0.15}$ & $67^{+27}_{-21}$ & 
$8.6\pm0.5$ & $0.21\pm0.06$ & 1021/1058  \\
6. PL+ABSORI+LINE & $2.57\pm0.06$ & 6.4$^{f}$ & $66\pm28$ & $2.7^{+29.2}_{-
1.5}\times10^{4}$ & $15.7^{+15.2}_{-6.2}$ & 1019/1059 \\
7. PL+PEXRIV+LINE & $2.68^{+0.07}_{-0.03}$ & 6.4$^{f}$ & $38^{+38}_{-29}$ & 
$2.4^{+12.8}_{-1.9}\times10^{3}$ & $0.7^{+0.6}_{-0.4}$ & 1016/1059 \\ 
\hline
\end{tabular}
\end{table*}

\subsection{The 2.5--20~keV spectrum}

We first examined the spectra in the 2.5--20~keV band, ignoring
for the time being the lower energy data. A simple spectral fit using a
model consisting of a power-law continuum modified by Galactic absorption
of $N_{H}=6.4 \times 10^{20}$~cm$^{-2}$ (Dickey \& Lockman 1990) gives
a steep slope but a poor overall 
fit to the data (see Table~1, model~1). In particular there are clear
deviations in the data--model residuals around 5--9~keV (see Fig.~1).

The addition of a narrow ($\sigma = 0.01$~keV) Gaussian line at 6.4~keV,
to represent K$\alpha$ emission from neutral iron (model~2), improved the fit
significantly ($>99.99$\% significance in an F-test). The best-fit
energy of the line is consistent with 6.4~keV but the width of this line
is not well constrained using these data. Even after the addition of an
iron line the residuals show a deficit at $\sim$8~keV, particularly in
the \xte\ data. In order to quantify these features we fitted
a range of simple models (see Table~1). Specifically, we tested models
with a line and an edge at energies expected from neutral iron (model~3),
from helium-like iron (model~4), and with the line and edge energies free
to vary (model~5). The He-like features provided a better fit than neutral
features, but the best fit values lie between these two extremes (see
Fig.~2). The important point is that the iron K-edge is at an energy
clearly above that for neutral iron. We have verified that the edge
energy is not an artefact of trying to fit a possible broad emission
feature with a narrow line, by repeating the fit with lines of increasing
width, up to $\sigma=0.5$~keV. The absorption edge energy and
optical depth remained consistent with the values given above for the
narrow line fit. In particular, the measured energy of the edge implies
an origin in strongly ionised material. This could, in principle, lie
along the line of sight (a `warm absorber') or arise by reflection
from optically thick matter having a highly ionised surface layer (an
`ionised reflector').

\begin{figure}
 \begin{center}
\rotatebox{-90}{\includegraphics[width=5.5cm]{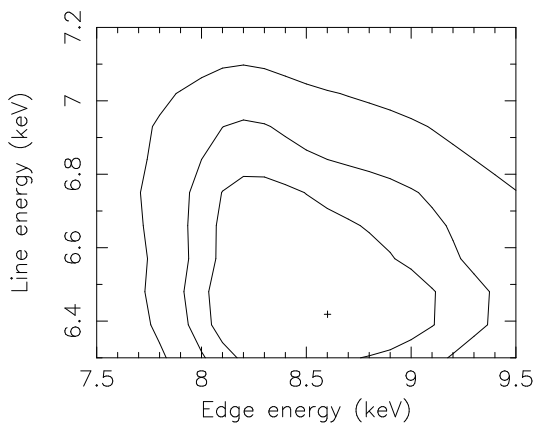}}
 \end{center}
 \caption{ Confidence contours for the iron K-line and -edge energies, 
from
model~5, Table~1. Contours mark the 68, 90 and 99\% confidence
levels. The edge energy is clearly not consistent with a neutral iron
edge at 7.1~keV.}
\end{figure}

To pursue these alternatives we first fitted the spectra above 2.5~keV
with the \textsc{absori} warm absorber model contained within the
\textsc{xspec}
package. The free parameters were power-law slope and
normalisation, iron line equivalent width, hydrogen column density 
and the ionisation parameter ($\xi= L / n r^{2}$; where
$n$ is the particle density at a distance $r$ from a source of ionising
luminosity $L$, measured in the 0.005--20~keV range). The
temperature of the absorbing gas was fixed at $10^{6}$~K. This model
provides a good fit to the data ($\chi^{2}_{\nu}=0.96$; see Table~1,
model~6). An unusually high $\xi$ is needed to explain the energy of the
iron edge, with a column density in excess of $10^{23}
\rm~cm^{-2}$. Such a highly ionised absorber, if indeed
stable, will be essentially transparent at lower energies.

As an alternative we fitted the 2.5--20~keV spectrum with the ionised 
reflector model \textsc{pexriv} in \textsc{xspec} (Magdziarz \& Zdziarski 
1995). This model consists of a power-law and reflection component from 
ionised material but does not contain the emission lines expected
from such a reflector. 
The energy of the iron line is only poorly constrained but 
is consistent with emission in the range 6.3--6.7~keV (see Fig.~2). 
We therefore added a narrow emission line at
6.4~keV to model the iron K$\alpha$ line.
This is not necessarily inconsistent with emission from an 
ionised disc as the line is expected to be broadened and redshifted 
if the emitting region is in close proximity to the central black hole.
The free parameters in the fit were the power law index and
normalisation, iron line equivalent width, the reflection strength ($R$)
and the ionisation parameter. The inclination of
the reflector was fixed at 30\degg (a value which is unlikely to be
exceeded in NLS1s) and the elemental abundances were assumed to be solar.
The surface temperature of the reflector was
fixed at $10^{6}$~K, consistent with values expected for an ionised disc
(e.g., Ross \et 1999). This model also provides a good fit to the data
($\chi^{2}_{\nu}=0.96$; Table~1, model~7), comparable with the warm
absorber model, with plausible values for both $\xi$ and $R$.

\begin{table*}
\centering
\caption{Results of simultaneous fits to the \asca\ and \xte\ data
in the 0.8--20~keV range. The columns give the following information:
(1) the models (as defined in the text);
(2) the power law photon index;
(3) the black body temperature (eV);
(4) the disk ionisation parameter ($\xi$);
(5) the strength of the reflection component ($R$); 
(6) the equivalent width of the iron K$_\alpha$ line (eV);
(7) the equivalent width of the O~\textsc{viii} recombination feature (eV);
(8) the temperature derived from the O~\textsc{viii} feature (eV);
(9) the best-fit $\chi^{2}$ and number of degrees of freedom.
Note that the model 4 fit includes \rosat\ data. } 
\begin{tabular}{@{}lcccccccc@{}}                 
Model & $\Gamma$ & kT & $\xi$ & R & EW$_{Fe}$ & EW$_{OVIII}$ &
kT$_{OVIII}$ & $\chi^{2}$/dof \\   
 (1)    &  (2)   & (3) & (4)  & (5) & (6)   & (7)    & (8) & (9) \\
\hline
1. PL+PEXRIV & 2.68$^{f}$ & -- & 2400$^{f}$ & 0.7$^{f}$ & 38$^{f}$ & --
& -- & 2040/1564 \\
2. PL+PEXRIV+BB & $2.66\pm0.03$ & $94^{+6}_{-8}$ & 
$1500^{+700}_{-600}$ & $0.6\pm0.2$ & $45\pm31$ & -- & -- &
1595/1557 \\
3. PL+PEXRIV+LINES & $2.69\pm0.03$ & -- & 1975$^{+345}_{-270}$ &
$0.9\pm0.2$ & $<47$ & $59\pm9$ & $50\pm20$ & 1596/1557 \\
4. PL+PEXRIV+LINES & $2.71\pm0.02$ & -- & 2000$^{f}$ & $0.94\pm0.08$ 
& $<45$ & $61\pm10$ & $45\pm12$ & 1619/1574 \\ 
\hline
\end{tabular}
\end{table*}

\subsection{The 0.8--20~keV spectrum}

Having obtained a satisfactory description of the hard X-ray
spectrum of Ark~564 we extended our analysis to cover the 0.8--20~keV
range, but exclude the SIS data below 1.0~keV in order to avoid the
calibration uncertainties mentioned above. Applying the 
ionised reflector model (with parameters as in Table~1, model~7) gave some
reduction in the `soft excess,' due to the enhanced
reflectivity of the ionised matter. However the fit (Table~2, model~1)
leaves a substantial excess flux which can be well modeled 
by the addition of a $\rm~kT \approx 100$~eV black body 
component (leading to $\Delta \chi^{2} > 445$; Table~2, model~2).

The steep underlying power law, enhanced by the high reflectivity of the 
ionised disc, lead to a smaller black body component than would
otherwise be the case. We note, furthermore, that a more complete ionised
reflector model would include significant soft X-ray emission. Reference 
to the disc reflection model recently published by Ross \et (1999), suggests 
that ionised oxygen features, in particular the recombination continuum 
of O~\textsc{viii} (above 0.87~keV) and  the
O~\textsc{viii} Ly-$\alpha$ emission line (at 0.65~keV), 
will be prominent for
values of the ionisation parameters similar to those derived for Ark~564.
Therefore, as an alternative to the black body component, we
refitted the ionised reflector model with the addition of 
O~\textsc{viii} recombination at 0.87 keV (using the 
\textsc{redge} model in \textsc{xspec}). The result (Table~2, model~3)
was a good fit (very similar, in statistical terms, to the
previous model) with a best fit temperature of $\sim 50$ eV
({\it i.e.} within a factor of two of our earlier assumption of a 
disk temperature of $10^{6}$ K). We conclude from this that, to first order, 
both the iron K features and the soft excess in the
\asca\ and \xte\ observation of Ark 564 can be explained in terms of
reflection
from highly ionised matter, presumably the putative accretion disc in this
NLS1. Specifically, the anomalous spectral form of Ark~564 below 2~keV
may have a natural explanation in terms of the combined effects of deep
O~\textsc{viii} and Si~\textsc{xiii} and \textsc{xiv} edges (present in
the \textsc{pexriv} reflection continuum) together with associated 
recombination continuum and line emission (see Fig.~3). 

One important consequence of our ionised reflector description for
Ark~564 is the need to reassess the strong `primary' soft emission 
component, thought to be a major feature of NLS1 (and generally
associated with internally generated emission from the accretion disc)
and extending up to
1~keV. Indeed when we included the \rosat\ spectrum considered by Brandt
\et (1994) in our model fitting (assuming that the spectral shape of
Ark~564 is constant but allowing the relative level of the spectrum to vary
from epoch to epoch) we found that the ionised reflector, now including 
O~\textsc{viii} recombination {\it and} Ly-$\alpha$ emission (the latter
with an equivalent width of $\sim 120$ eV), allowed a reasonable fit down 
to 0.2 keV (Table 2, model 4).

\begin{figure}
 \begin{center}
\rotatebox{-90}{\includegraphics[width=5.5cm]{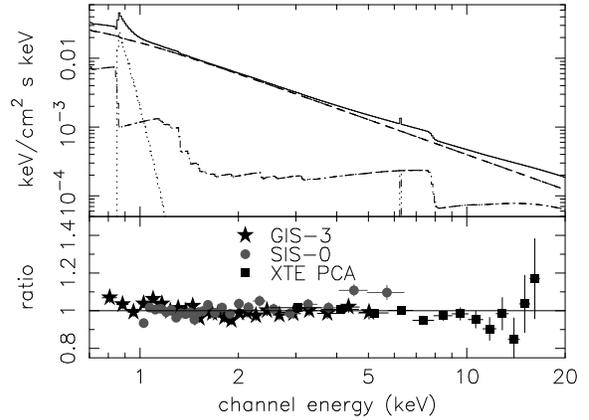}}
 \end{center}
 \caption{ Input model and data/model ratio from the \asca\ and \xte\ 
spectrum of Ark~564 (see model 3, table 2). The model spectrum (solid curve) 
represents the best-fit power law plus ionised reflector (dashed curves), 
including an O~\textsc{viii} recombination feature at 0.87~keV and iron line 
at 6.4~keV (dotted curves). The data/model 
residuals indicate a good fit, both in the iron K band and at
softer energies.} 
\end{figure}

Comastri \et (1998b) present a 
preliminary analysis of a \sax\ observation of Ark~564. 
Our interpretation of the X-ray spectrum of Ark~564 in terms of an
ionised reflector is broadly consistent with the measured \sax\ spectrum.

\begin{figure}
 \begin{center}
\rotatebox{-90}{\includegraphics[width=5.5cm]{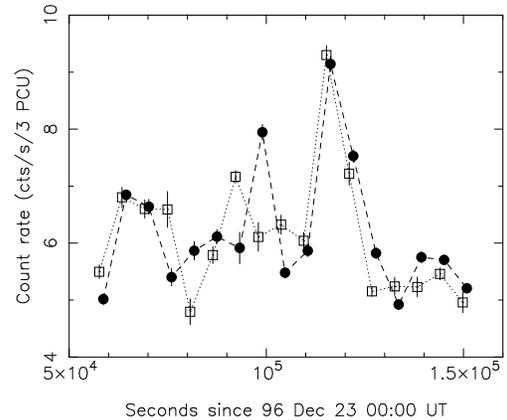}}
 \end{center}
 \caption{
Light curves for soft 0.5--2~keV \asca\ (open squares, dotted lines)
and 3--12~keV \xte\ (solid circles, dashed lines) data in orbital bins.
The \asca\ count rates have been scaled up by a factor of 4 for
comparison with the \xte\ data.
Note the strong ($\gs$50\%) flare near $ T = 1.15 \times 10^5 $~sec,
which is virtually identical in phase and amplitude in both bands. 
Errors are $\pm 1 \sigma$. }
\end{figure}

\section{ Temporal analysis }

These data were also used to study the variability properties of Ark~564.
Fig.~4 shows that the \asca\ soft (0.5--2~keV) and \xte\ hard (3--12~keV) band 
light curves reveal strong flaring, with an increase of $>$50\% in
a single orbit that is almost identical in amplitude and phase in both bands.

In order to quantify the variability in Ark~564, we computed the `excess 
variance' ($\sigma^2_{xs}$) in the same fashion as Nandra \et (1997),
except that paper used 128s bins while we use orbital bins.
For Ark~564, the hard and soft band excess variances were measured to be 
$ \sigma_{xs}^{2} = 0.0322 $ and 0.0325, respectively.
This compares with a typical value found by Nandra \et (1997) for `normal' 
Seyfert~1s with a similar luminosity ($ \sim 6 \times  10^{43} $~erg/s) of 
$ \sigma_{xs}^{2} = 0.005 $ in the \asca\ 0.5--10~keV band.

It is also interesting that Ark~564 shows almost identical variability 
amplitudes in the hard \xte\ and soft \asca\ bands.
Furthermore, the hard and soft band variations track almost perfectly, with no 
measurable lag, especially during the large flare. In terms of our present
model of Ark 564, in which a substantial fraction of the soft X-ray flux
is reprocessed harder radiation, the
observed limit on the lag ($_{\sim}^{<}$ 96 min), gives a maximum size for
the effective reprocessing region of $ \sim 2 \times 10^{14} $~cm. 

\section{Discussion}
 
The simultaneous \asca\ and \xte\ data presented here give the first
determination of the spectrum of an `ultra-soft' NLS1 extending above
10~keV. The spectrum is extremely steep ($\Gamma \approx 2.7$) and shows
little sign of flattening at harder energies. There is good evidence
for both K-edge and line features from ionised iron 
and the spectrum also shows a strong excess over the best-fit 
power law below 1.5~keV. Due to calibration problems with the 
\asca\ detectors at low
energies (see Iwasawa, Fabian \& Nandra 1999), the data have not
been fitted below 0.8~keV and so the form of this soft excess is not well
determined by these data.

The existence of strong soft X-ray emission and an underlying power
law much steeper than typical of BLS1s and quasars are well established
characteristics of the X-ray spectra of NLS1s. It has been suggested that
the two features are linked, whereby Compton cooling of the hard X-ray
source (possibly in a disc corona), and hence its steeper power law, is a
consequence of the strong soft EUV flux (Pounds \et 1995). In this
picture, the soft component, probably peaking in the hidden EUV band, is
intrinsic emission from the accretion disc, which is expected to be
stronger in high accretion rate objects (e.g. Szuszkiewicz \et 1996).

In the present analysis we now find other spectral features, namely an
ionised iron K-edge and recombination emission below $\sim$1.5~keV,
indicative of reflection from an ionised disc. The higher
level of irradiation thought to occur in NLS1s would lead naturally to the
surface layers of the disc becoming strongly ionised (Matt \et 1993). In
contrast, the alternative interpretation we have considered for the
observed
iron K-edge, in terms of absorption in a large column of highly
ionised gas, is
less attractive, since it may well be unstable (Netzer 1996) and because
such material would be transparent in the soft X-ray band, leaving
the need for a separate explanation of the the observed
spectral
features below $\sim$2~keV.

An important consequence of the steep underlying power law of Ark~564,
enhanced at low energies both by the high reflectivity of the disc and
the additional line and recombination emission, is that the need for a primary 
emission `soft excess,' at least within the observable X-ray band, must be
reconsidered when better data from e.g., \xmm\ and {\it Chandra} become 
available.
However, the circumstantial evidence for the soft X-ray/EUV flux to dominate
energetically in NLS1 remains persuasive, particularly in providing
a natural explanation for the steep hard power law spectrum and the
absence of broad optical lines.

Finally we note that evidence for reflection from an ionised disc has
been reported in several GBHC, exhibiting
strong, ionised edges with weak, Compton-broadened Fe lines (e.g., Zycki
\et 1997, 1998; but see Done \& Zycki 1999). The similarity of NLS1 to GBHC in
their high flux mode has been noted earlier (Pounds \et
1995). It is interesting to speculate that the reported weakness in
higher luminosity AGN (e.g. Reeves \et 1997), of the reflection
components commonly seen in BLS1 (George \et 1998), may be a due to the
accretion disc material in the former being highly ionised, with the classical
indicators of cold reflection, namely strong iron fluorescence and a
continuum hump near 10~keV, consequently reduced.

\section*{Acknowledgments}
This research made use of data obtained from the LEDAS and HEASARC. 
We thank Andy Young for
providing data files on the Ross \et ionised disc model. 
SV acknowledges a research studentship from PPARC.
RE acknowledges support from NASA grant NAG~5-3295.

\bsp
\label{lastpage}
\end{document}